# Ultra-small topological spin textures with size of 1.3nm at above room temperature in $Fe_{78}Si_9B_{13}$ amorphous alloy

Weiwei Wu[1,2#], Huaping Zhang[1,3#], Hong Wang[1#], Chao Chang[1,2], Hongyu Jiang[1], Jinfeng Li[1,4], Zhichao Lv[3], Laiquan Shen[1,3], Hanqiu Jiang[5,6], Chunyong He[5,6], Yubin Ke[5,6], Yuhua Su[7], Kosuke Hiroi[7], Zhendong Fu[3], Zi-An Li[1,2]\*, Lin Gu[1,2], Maozhi Li[8], Dong Ma[3], and Haiyang Bai[1,2,3,4]\*

[1]*Institute of Physics, Chinese Academy of Sciences, Beijing 100190, China*

[2]*School of Physical Sciences, University of Chinese Academy of Sciences, Beijing 101408, China*

[3]*Songshan Lake Materials Laboratory, Dongguan, Guangdong 523808, China*

[4]*College of Materials Science and Opto-Electronic Technology, University of Chinese Academy of Sciences, Beijing 100049, P. R. of China*

[5]*Institute of High Energy Physics, Chinese Academy of Sciences (CAS), Beijing, 100049, China*

[6]*Spallation Neutron Source Science Center, Dongguan, 523803, China*

[7]*J-PARC Center, Japan Atomic Energy Agency, Tokai, Ibaraki, 319-1195, Japan*

[8]*Department of Physics, Renmin University of China, Beijing 100872, China*

[#]These authors contributed equally: Weiwei Wu, Huaping Zhang and Hong Wang

\*Corresponding author. Email: zali79@iphy.ac.cn, hybai@iphy.ac.cn

## Abstract

**Topologically protected spin textures, such as skyrmions[1,2] and vortices[3,4], are robust against perturbations, serving as the building blocks for a range of topological devices[5-9]. In order to implement these topological devices, it is necessary to find ultra-small topological spin textures at room temperature,**

because small size implies the higher topological charge density, stronger signal of topological transport[10,11] and the higher memory density or integration for topological quantum devices[5-9]. However, finding ultra-small topological spin textures at high temperatures is still a great challenge up to now. Here we find ultra-small topological spin textures in $Fe_{78}Si_9B_{13}$ amorphous alloy. We measured a large topological Hall effect (THE) up to above room temperature, indicating the existence of highly densed and ultra-small topological spin textures in the samples. Further measurements by small-angle neutron scattering (SANS) reveal that the average size of ultra-small magnetic texture is around 1.3nm. Our Monte Carlo simulations show that such ultra-small spin texture is topologically equivalent to skyrmions, which originate from competing frustration and Dzyaloshinskii–Moriya interaction[12,13] coming from amorphous structure[14-17]. Taking a single topological spin texture as one bit and ignoring the distance between them, we evaluated the ideal memory density of $Fe_{78}Si_9B_{13}$, which reaches up to $4.44 \times 10^4$ gigabits (43.4 TB) per $in^2$ and is 2 times of the value of $GdRu_2Si_2$[18] at 5K. More important, such high memory density can be obtained at above room temperature, which is 4 orders of magnitude larger than the value of other materials at the same temperature. These findings provide a unique candidate for magnetic memory devices with ultra-high density.

Topological magnetic materials contain particle-like topological spin textures such as skyrmions[1,2], merons[19] and hopfions[20], providing a promising route toward future information technologies, including the nonvolatile magnetic memory[6], non-von Neumann computing architecture for unconventional computing paradigms[9], artificial synapse device in neuromorphic systems with adaptive learning function[8], and topological quantum computation with Majorana bound states in proximity effect of magnetic skyrmions and s-wave superconductors[7,5]. However, to find the ideal topological magnetic materials, two physical limits must be broken simultaneously, namely the small-size limit and the high-temperature limit[9]. The former is to search topological spin textures with extremely small sizes which is urgently-needed for the miniaturized devices with high-density information carriers[6,9]. While the later requires topological spin textures to be stable at above room temperature to achieve a wide range of practical applications[6]. So far, extensive efforts have been devoted to this issue, with unsatisfactory results. Although small-size skyrmions of ~1.9 nm typical size have been observed, the measured temperature is as low as 5 K[18]. At above room temperature, the reported sizes of skyrmions are all around several tens to several hundred nanometers. Thus, there has been no convincing evidence for the existence of ideal topological magnetic materials that simultaneously break through the both limits. Here, we provide an evidence for soft-magnetic $Fe_{78}Si_9B_{13}$ amorphous alloy by revealing the existence of ultra-small topological spin textures with size of ~1.3nm at above room temperature. As an estimate, it could provide the highest topological memory density at above room temperature, reaching up to $4.44\times10^4$ gigabits (43.4 TB) per $in^2$. This is at least 4 orders of magnitude larger than that in the other topological magnetic materials at the same temperature. Meanwhile, the $Fe_{78}Si_9B_{13}$ amorphous alloy has excellent soft magnetic properties with low energy consumption, and unique advantages in the high-speed/frequency and high-efficiency magnetic devices. Therefore, our findings provide a unique candidate for topological memory devices working at room temperature with ultra-high density, high speed and high efficiency.

Due to the complexity of disorder, previous studies focused on topological spin textures usually avoid considering amorphous materials due the fact that it is difficult for the classical condensed physics to deal with disorder systems, specially on the analysis of band theory, and many possibilities presented by amorphous materials have been missed. Here, we specially focus on the topological spin textures in an amorphous $Fe_{78}Si_9B_{13}$ alloy. The reasons for choosing $Fe_{78}Si_9B_{13}$ metallic glasses (MGs) are in many aspects. Firstly, due to the structure disorders, magnetic MGs can accommodate various types of interactions, particularly, the non-collinear spin texture (Fig. 1A) described by a 'combed hair' model[21] was previously confirmed in $Fe_{78}Si_9B_{13}$ by neutron scattering and Mössbauer spectra[22-24], which originated from the competing frustration interactions. Meanwhile, according to A. Fert[14], a non-vanishing Dzyaloshinskii–Moriya interaction (DMI)[12,13] should work in amorphous materials due to inversion symmetry breaking inherently. And this non-vanishing DMI has been recognized to responsible for the skyrmions found in amorphous $FeGe$[17] and $DyCo_3$[15,16] films. Secondly, $Fe_{78}Si_9B_{13}$ MGs have a high curie temperature $T_c$ (683K), and the composition can vary continuously over a wide range, i.e., the magnetic interactions can be tuned continuously at a wide range. Moreover, iron-based MGs such as $Fe_{78}Si_9B_{13}$ are mature, cheap and green commercial-available products. Therefore, the $Fe_{78}Si_9B_{13}$ MGs can serve as an ideal magnetic system for investigation of topological spin textures at above room temperature. The amorphous nature of the sample is verified by XRD, HRTEM and selected area electron diffraction patterns (see Fig. 1B).

The initial clues of the existence of topological spin textures in $Fe_{78}Si_9B_{13}$ MG come from the measurements of transport properties. In the adiabatic limit, the topological spin textures would generate real-space Berry curvature, i.e., an effective magnetic field, acting on the conduction electrons, inducing a transverse voltage drop known as topological Hall effect (THE). Therefore, THE itself provides a powerful probe to detect unconventional magnetic structures. Generally, if THE is nonzero, the Hall resistivity ($\rho_{xy}$) has three contributions as below:

$$\rho_{xy} = \rho_{OHE} + \rho_{AHE} + \rho_{THE} = R_0 H + R_s M_z + \rho_{xy}^T. \qquad (1)$$

Here, $R_0$ is the ordinary Hall coefficient, $R_s$ is the anomalous Hall coefficient, $M_z$ is the magnetization, and $\rho_{xy}^T$ is the topological part. Figure 1c and 1d present the measured $H$-dependence of $\rho_{xy}$ and $M_z$ at different temperatures ($T$), respectively. Although, both of $\rho_{xy}$ and $M_z$ saturate at the high field region, they show distinct behavior at low field region. As can be seen, magnetization process has two procedures with distinct slopes: a first steep magnetization at low field followed by a relative mild linear magnetization behavior until saturation[25,26]. Since $\rho_{xy}^T$ is zero above saturation magnetic field, both of $R_0$ and $R_s$ can be obtained via linear fitting of $\rho_{xy}$-$H$ curves at the high field region. After that, the topological part $\rho_{xy}^T$ can be obtained by subtracting $R_0 H$ and $R_s M_z$ from $\rho_{xy}$.

Figure 2a shows details of this procedure at 350K. Using this way, the $H$ dependence of $\rho_{xy}^T$ at different temperatures can be obtained. As shown in Fig. 2b, the THE signal is noticeable and in the magnitude of μΩcm, existing in the whole measurable temperature range. There are two peak regions with opposite signs, including a deep and large negative valley at lower fields, following a relative positive hump at higher fields. The maximum absolute value of $\rho_{xy}^T$ at lower field and higher field respectively is plotted as a function of temperature in Fig. 2c. The maximum value is 1.8 μΩcm at 350K which is much larger than the value of B20 systems[27-29]. Taking THE data for all temperatures together, the ($H$-$T$) phase diagram is displayed in Fig. 2d. More details of THE are discussed in Supplementary Information A. The measured THE means the existence of abundant topological spin textures in the $Fe_{78}Si_9B_{13}$ MG.

Theoretically, THE reflects the real space Berry phase produced by the topological spin textures such as merons or skyrmions, which can be quantified by the topological charge (TC) as[11]:

$$Q = \frac{1}{4\pi} \int d^2r \mathbf{n} \cdot \left( \partial_x \mathbf{n} \times \partial_y \mathbf{n} \right), \qquad (2)$$

where $\mathbf{n}$ is a unit vector describing the local spin direction. Based on strong ferromagnetic exchange coupling model, the THE is determined by[10,30]:

$$\rho_{xy}^T = P R_0 n_Q \Phi_0, \qquad (3)$$

where $P$ is spin polarization, $R_0$ is the ordinary Hall coefficient, $n_Q$ is areal density of TC and $\Phi_0 = h/e$ is flux quantum. This equation provides a feasible way to investigate the underlying topological magnetic structures in $Fe_{78}Si_9B_{13}$ from the transport measurements. Using the assumption of 100% spin polarization and the maximum absolute value of $\rho_{xy}^T$ at 300K, 1.6 μΩcm, the areal density of TC, $n_Q$, is evaluated as 0.012/nm² from equation (3). The high areal density here corresponds to the small size of topological spin texture[30].

Next, to determine the size of topological spin texture in $Fe_{78}Si_9B_{13}$ MG, we have brought in the small-angle neutron scattering (SANS), a powerful tool wildly used to detect nanoscale (1-100 nm) magnetic structures in bulk materials. Figure 3a shows the SANS intensity of magnetic contribution $I_{MAG}$ (see Supplementary Information E for more details) at different momentum transfer $Q$ in $Fe_{78}Si_9B_{13}$ MG at room temperature. Whereas the $I_{MAG}(Q)$ decays much faster at smaller $Q$ regions, it would decay slower at larger $Q$ regions. According to R. García et al.[31] and C. Bellouard et al.[32], if the sizes of magnetic textures are polydisperse in the sample, there may be no obvious peak in the $I_{MAG}(Q)$ curve. Therefore, to extract the size information of magnetic texture, we used the model proposed by R. García et al.[31] to fit our results. This model describes the contributions of two kinds of magnetic textures with different sizes to the intensity of $I_{MAG}$. One is the smaller spin texture, and the other is the larger-size magnetic structure in domain walls. Following García's model, the $I_{MAG}(Q)$ is fitted with the equation as:

$$I_{MAG}(Q) = \Delta\rho_{MAG}^2 \sin^2\alpha \left[ f_{p1} \frac{\int_0^\infty g_1(R) V_1^2(R) F^2(Q,R) dR}{\int_0^\infty g_1(R) V_1(R) dR} + f_{p2} \frac{\int_0^\infty g_2(R) V_2^2(R) F^2(Q,R) dR}{\int_0^\infty g_2(R) V_2(R) dR} \right], \quad (4)$$

where $\Delta\rho_{MAG}$ is the magnetic contrast of the textures with respect to the matrix, $\alpha$ is the angle between the scattering and the texture-magnetization directions, $V(R)$ and $F(Q,$ $R$) are the volume and scattering factor of a spin texture with radius $R$, respectively, $g(R)$ is the log-normal distribution of the radius, $f_p$ is the volume fraction for each kind of spin texture. As shown in Fig. 3a, the $I_{MAG}(Q)$ can be well described by the equation over the entire $Q$ range. Figure 3b shows the two optimized size distributions of spin textures. Whereas one distribution has an average diameter as small as around 1.3 nm, the other one has a lager average diameter about 47 nm. More importantly, the volume fraction of small spin texture in the sample is about 89%, while that of large size texture is about 11%. For the larger texture, Lorentz transmission electron microscopy (LTEM) images which is discussed in the Supplementary Information B show that its size is about 50nm, and it exists on the Bloch line of the magnetic domain walls and the overall proportion in the matrix is sparse. These observations are also consistent with the results of SANS. (See Supplementary Information E for more details). The above results provide strong evidence that even at room temperature there are plenty of ultra-small topological spin textures, whose diameter is down to 1.3 nm in average.

In the following part, the properties and potential applications are compared for different topological magnetic materials containing topological spin textures. In Fig. 3c, we compared these topological magnetic materials from three aspects, including the maximum THE, the measured temperature and the size of the topological spin textures. The maximal value of THE at above room temperature in our sample is 1.8 μΩcm, which is much larger than that in B20 chiral magnets[27-29,34] and reaches the same order of THE in frustrated triangular-lattice $Gd_2PdSi_3$[43]. We expect the measurement at high temperatures would give a larger THE due to the thermal excitation. More importantly, the size of the topological spin texture responsible for THE in $Fe_{78}Si_9B_{13}$ is about 1.3nm in average. To the best of our knowledge, it is so far the smallest at room temperature in all reported systems[45]. Then, taking one topological texture as one bit and ignoring the distance between textures, we can evaluate the ideal memory density for the several materials. As shown in the inset of Fig. 3c, the $Fe_{78}Si_9B_{13}$ has the highest memory density, reaching up to $4.44 \times 10^4$ gigabits (43.4 TB) per $in^2$ which is 2 times of the value of $GdRu_2Si_2$ at 5K. More importantly, such high memory density can be obtained at above room temperature, which is 4 orders of magnitude larger than the value of other materials at the same temperature. Therefore, our findings provide a unique candidate for magnetic memory devices with ultra-high density.

These results of transport and neutron measurements indicate that $Fe_{78}Si_9B_{13}$ MG belongs to ideal topological magnetic materials with ultra-small topological spin texture at room temperature. Next, we investigate the forming mechanism of this ultra-small topological spin textures. In fact, amorphous $Fe_{78}Si_9B_{13}$ alloy contains various complex magnetic interactions, including the DMI[12,13] and exchange frustrations[21,22]. Therefore, the Monte Carlo (MC) simulations were conducted with the following Hamiltonian:

$$H = \sum_{\langle i,j \rangle} \left[ -J_1 \left( \bm{S}_i \cdot \bm{S}_j \right) + \bm{D}_{ij} \cdot \left( \bm{S}_i \times \bm{S}_j \right) \right] + \sum_{\langle\langle i,k \rangle\rangle} J_2 \left( \bm{S}_i \cdot \bm{S}_j \right) - B_z \sum_i S_{i,z}$$

where $\bm{S}_i$ is the spin on site $i$, $\langle i,j \rangle$ and $\langle\langle i,k \rangle\rangle$ means the nearest neighbor (NN) and next-nearest neighbor (NNN), respectively. In our simulation, $|\bm{S}_i| = 1$ and a 2D amorphous system was employed. $J_1$ (=1.0) is the NN ferromagnetic coupling, and $J_2$ is the NNN antiferromagnetic coupling, while $\bm{D}_{ij} = D\hat{\bm{r}}_{ij}$ is the vector of the DMI between NN sites $i$ and $j$. Firstly, two extreme situations were studied and compared, including model A without DMI ($J_2$=0.5, $D$=0) and model B without exchange frustrations ($J_2$=0.0, $D$=0.3). As shown in Fig. 4, model A and B show significant differences in both the magnetization curves and the topological charges. For model A, the magnetization increases continuously and smoothly with increasing external field. Whereas for model B, M shows typically two-step behaviors, which increases fast at first and then stagnates. The differences become more remarkable when considering the topological charge. As shown in Fig. 4e, f, the average topological charge is always zero for model A. Whereas for model B, there exist nonzero topological charges, responsible for the THE. This result infers that DMI plays an important role in producing THE in glassy materials. Secondly, we also conducted MC simulations for mixed situations by adding part of exchange frustration to model B. As shown in Fig. 4g, f, adding small exchange frustration ($J_2$=0.1, $D$=0.3) can effectively promote the density of topological charge. Figure 4i-m show the typical spin configurations of these vortices for models with different interactions, and these spin textures are topologically equivalent to antiskyrmions or skyrmions. In Fig. 4i, both skyrmions and antiskyrmions can be stabilized by the exchange frustration without DMI. While the skyrmions are preferred and stabilized when combining exchange frustration and DMI, as shown in Fig. 4j-m. More importantly, direct studies on the magnetic structures also show that the size of vortex decreases dramatically as the increase of exchange frustration $J_2$. Thus, the large THE observed in amorphous $Fe_{78}Si_9B_{13}$ alloy could be the combined effects of DMI and exchange frustration i.e., the DMI produces nonzero topological charges, the exchange frustration reduces the vortex size and enhances the density.

In summary, we have found a kind of topological magnetic material with 1.3 nm topological spin textures at above room temperature, which provides a unique candidate material for magnetic memory devices with ultra-high density. Our findings indicate that a large class of non-collinear amorphous magnets discovered for a long time are topological magnets, which have abundant nontrivial spin structures with different topological configurations and pave the way for a series of practical applications of topological devices.

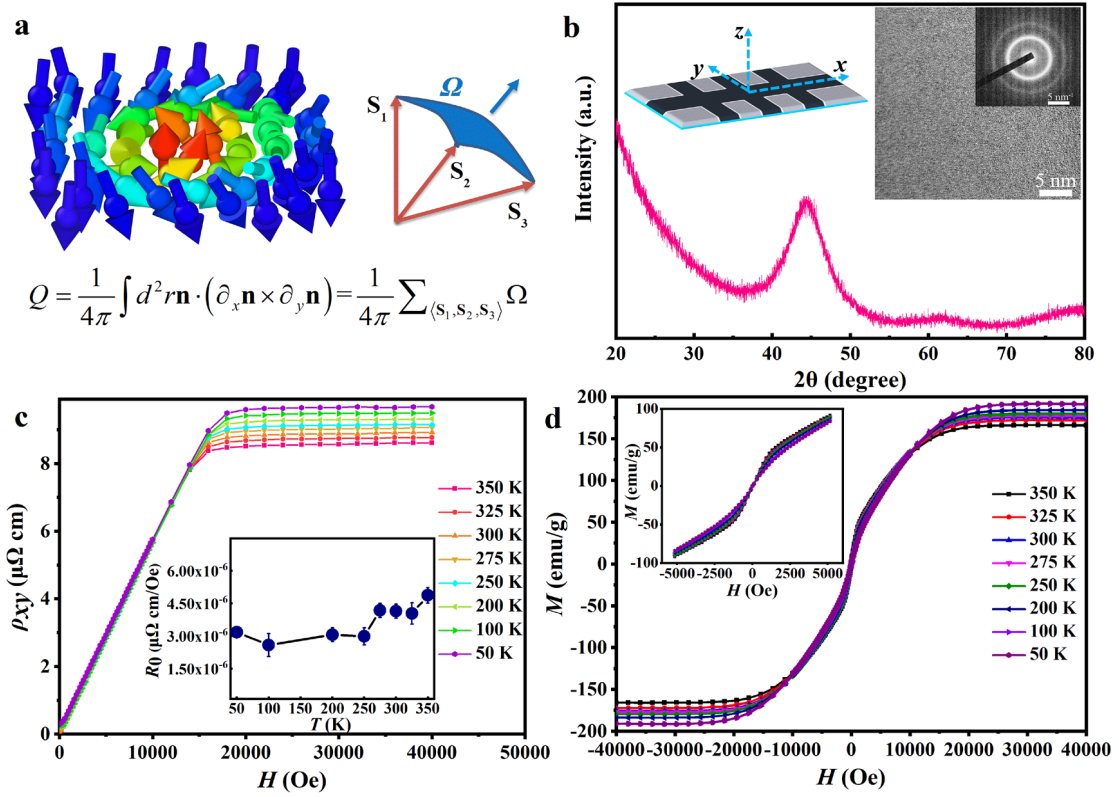

**Figure 1. a**, a schematic of non-collinear spin texture and spin chirality consisted by three non-coplanar spins. $S_1$, $S_2$ and $S_3$ are three neighbouring spins (brown arrows). $\Omega$ is a solid angle subtend by these three non-coplanar spins and induces an effective magnetic field (blue arrow) in the adiabatic limit. **b**, XRD curve of $Fe_{78}Si_9B_{13}$ (pink line). Inset of the left panel, a schematic Hall bar of the MG ribbon cut by laser beam. Inset of the right panel, HRTEM images and the selected area electron diffraction pattern. **c**, Hall effect at different temperatures. Inset, temperature dependence of ordinary Hall coefficient $R_0$. Error bars are the standard deviations of the linear fits when determining the AHE signal at high field. **d**, Magnetization $M$ versus out-of-plane magnetic field $H$ at different temperatures. Inset shows details of M-H curves at low fields.

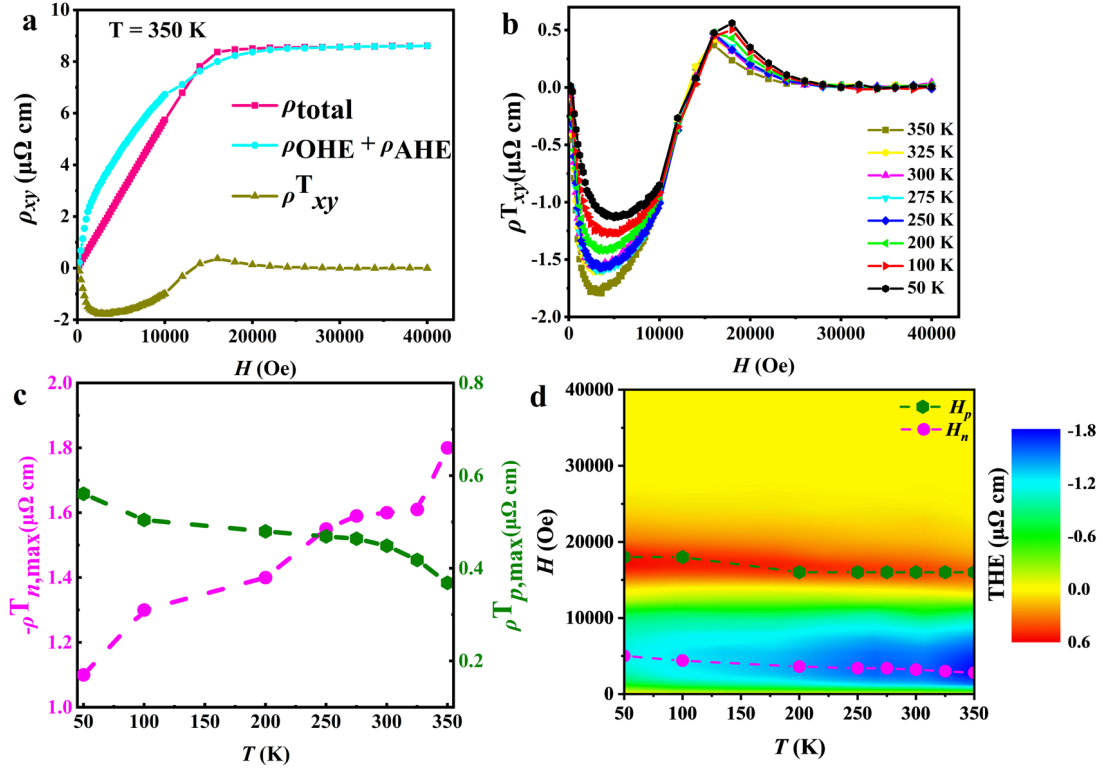

**Figure 2. a**, Decomposition procedure extracts $\rho_{xy}^T$ at 350K from the total Hall signal by $R_0H + R_sM_z$. **b**, $\rho_{xy}^T$ versus magnetic field $H$ at different temperatures. **c**, Temperature dependence of absolute value of maximum $\rho_{xy}^T$ at low fields and high fields respectively, $-\rho_{n,\max}^T$ and $\rho_{p,\max}^T$. The dash line is guide for eyes. **d**, the ($H$-$T$) phase diagram of THE. The position of THE maximum at different temperatures are shown as $H_n$ and $H_p$, respectively. (*n* and *p* represent negative and positive respectively)

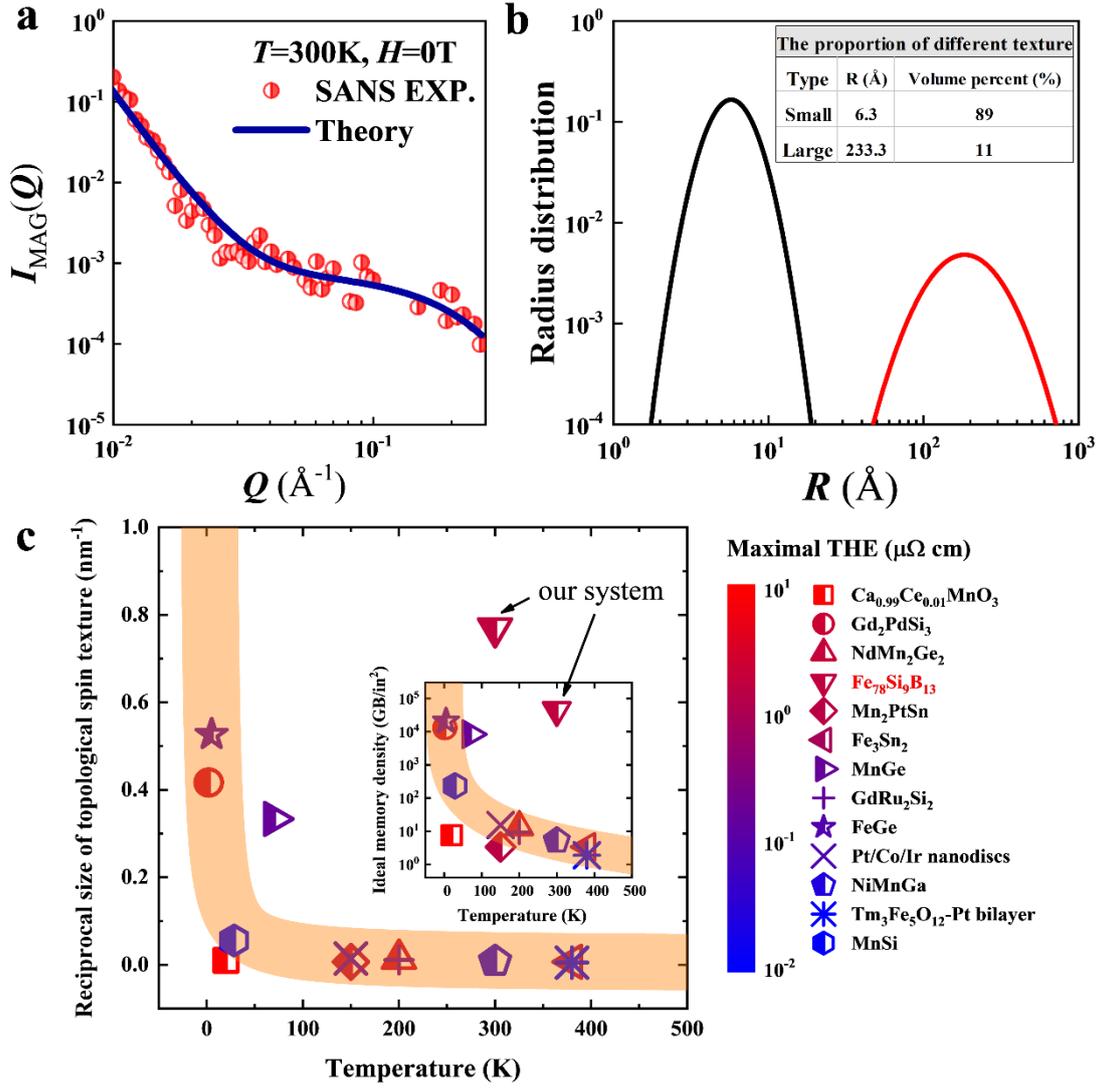

**Figure 3. a**, SANS intensity of magnetic contribution $I_{MAG}(Q)$ and the best fits (blue solid curves) based on Eq. (4). **b**, Log-normal distributions of the texture sizes corresponding to the best fits. The inset shows the proportion of different texture. **c**, the comparison of different materials hosting topological spin textures. The reciprocal size of topological spin texture is used to emphasize the topological texture with small size. The maximal THE and the corresponding temperature are used to compared $Fe_{78}Si_9B_{13}$ to other magnetic materials[18,27,28,33-44]. The inset shows the ideal memory density of the various materials. Detailed information is given in Supplementary Information Table 1.

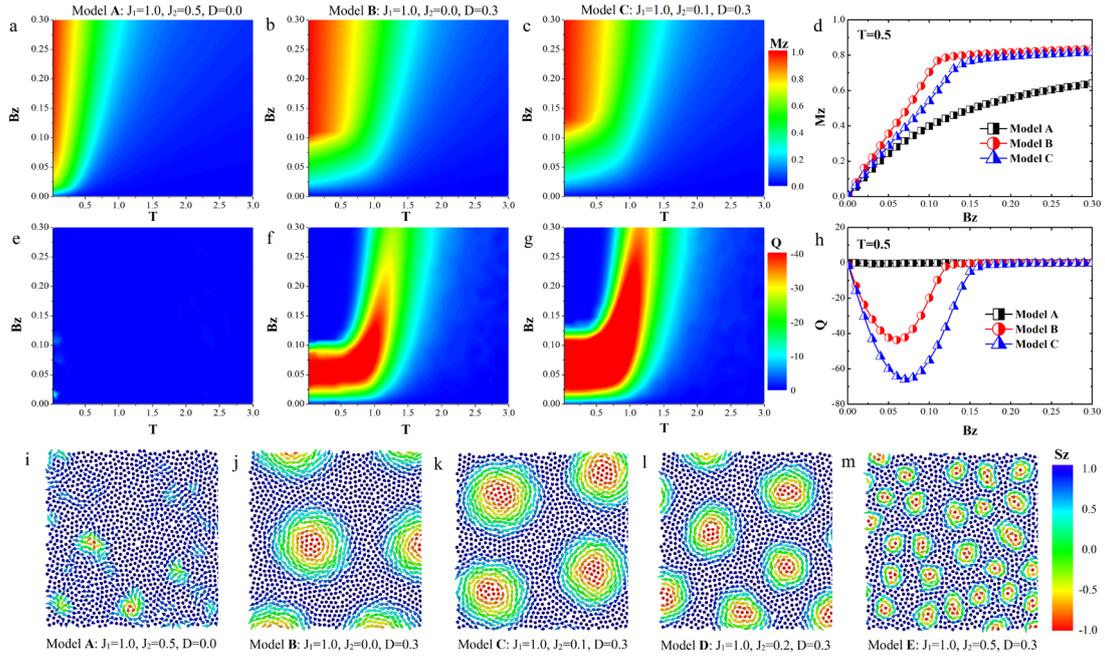

**Figure 4. a-c**, The count plots of the magnetization $M$ in $B_z$-$T$ phase plane for model A ($J_2 = 0.5$, $D = 0.0$), model B ($J_2 = 0.0$, $D = 0.3$), and model C ($J_2 = 0.1$, $D = 0.3$), respectively. **d**, the magnetization curves at $T=0.5$ for the three models. **e-g**, The count plots of topological charge $Q$ for the three models. **h**, The $Q$ curves at $T=0.5$ for the three models. **i**, The typical spin configuration of vortex structures for model A ($J_2 = 0.5$, $D = 0.0$). **j-l**, The typical spin configurations of vortex structures for models with different exchange frustration: **j** for model B ($J_2 = 0.0$, D=0.3), **k** for model C ($J_2 = 0.1$, D=0.3), **i** for model D ($J_2 = 0.2$, D=0.3), **m** for model E ($J_2 = 0.5$, D=0.3). While the simulated system contains a total of 25600 spins (L=171), the plotted region here contains only small part of them (L=40). Besides, for all the studied models, the NN ferromagnetic coupling ($J_1$=1.0) remains the same. More information about the simulations and analysis can be found in the Supplementary Information G.

## Methods

**Sample preparation and characterization.** The commercially available $Fe_{78}Si_9B_{13}$ MG ribbons with a uniform thickness of 25 μm were fabricated by single-roll melt-spinning method. The X-ray diffraction (XRD) with Cu $K_α$ radiation and the High-resolution transmission electron microscopy (HRTEM) were used to verify the amorphous nature of the samples. The TEM samples were carefully prepared by ion milling with 2-keV argon ions at liquid-nitrogen temperature and the HRTEM observations were conducted using a JEOL-2100 TEM.

**Magnetic and electrical transport property measurements.** The magnetization was measured by a vibrating sample magnetometer (VSM). To characterize the electrical transport properties, we used the laser beam to cut MG ribbon into a standard Hall bar. Electric contacts were fabricated with platinum wire and silver paint. The Hall resistivity ($ρ_{xy}$) was measured by a physical property measurement system (PPMS, Quantum Design). The thickness of our sample is 25μm, which is negligible compared to length and width (several millimeters). The demagnetization factor $N_d ≈ 1$ when the magnetic field is perpendicular to the plane of the ribbon.

**LTEM measurements.** LTEM were carried out using a JEOL 2100F microscope combined with heating holder (Gatan Inc. Model 625.TA) and cooling holder of nitrogen type (Gatan Inc. Model 626TA). Quantitative magnetization phase contrast was analyzed using transport of intensity equation (TIE)[46] and method using homemade code. The detailed LTEM images and analysis can be referred in Supplementary Information B.

**Small angle neutron scattering (SANS) measurements.** The ribbon was cut into 12 × 12 mm² pieces, which were stacked up and glued on the corners to make a 12 × 12 × 0.2 mm³ sample for SANS measurements. SANS experiments were carried out on the time-of-flight small-and-wide-angle neutron scattering instrument TAIKAN (BL15) at Materials and Life Science Experimental Facility (MLF) in Japan Proton Accelerator Research Complex (J-PARC)[47].

**SANS data analysis.** Generally, the total SANS intensity comprises nuclear and magnetic contributions: $I_{TOT}(Q, \alpha)=I_{NUC}(Q)+I_{MAG}(Q)\sin^2\alpha$, where $Q$ is the scattering vector and $\alpha$ is the angle between $Q$ and magnetization vectors. When $H$ is large enough to saturate the sample and parallel to $Q$, $\alpha=0°$, and hence $I_{TOT}=I_{NUC}=I_{P-SAT}$. If $I_{TOT}$ is measured in a direction vertical to that of $H$, $\alpha=90°$ and hence $I_V(I_{TOT})=I_{NUC}+I_{MAG}=I_{P-SAT}+I_{MAG}$. Therefore, $I_{MAG-0T}(Q)= I_{V-0T}(Q) − I_{P-1T}(Q)$, in our sample. According to R. García et al.[31] and C. Bellouard et al.[32], if the magnetic texture is polydisperse in the sample, there may be no obvious peak in the $I_{MAG}(Q)$. Therefore, to extract the size information of magnetic texture, we used the model proposed by R. García et al.[31] to fit our results. This model describes the contribution of two magnetic textures with different sizes to the intensity of $I_{MAG}(Q)$. The scattering cross section of single magnetic texture can be written as:[32]

$$\left(\frac{d\sigma}{d\Omega}\right) = N_p V_p^2 \Delta\rho_{MAG}^2 \sin^2\alpha F_p^2(Q) S(Q),$$

where $N_p$ is the number of magnetic textures in the sample, $V_p$ is the volume of these textures, $F_p(Q)$ is their form factor, $\Delta\rho_{MAG}$ is the magnetic contrast of the textures with respect to the sample, $\alpha$ is the angle between the scattering and the texture-magnetization directions, and $S(Q)$ is the structure factor corresponding to the textures. Due to the polydispersity of these magnetic textures in the sample, $S(Q)$ is set as 1 for the whole range of $Q$ values in the present data. Assuming that the magnetic texture is a spherical cluster with radius $R$, its form factor $F_p(Q)$ (normalized to 1) is given by the following expression:

$$F_p(Q) = 3\frac{\sin(QR) - QR\cos(QR)}{Q^3 R^3}.$$

Assuming the log-normal distributions of texture sizes, the total scattering cross section of two different average sizes can be written as:

$$\frac{1}{V}\left(\frac{d\sigma}{d\Omega}\right) = \Delta\rho_{MAG}^2 \sin^2\alpha \left[ f_{p1}\frac{\int_0^\infty g_1(R)V_1^2(R)F^2(Q,R)dR}{\int_0^\infty g_1(R)V_1(R)dR} + f_{p2}\frac{\int_0^\infty g_2(R)V_2^2(R)F^2(Q,R)dR}{\int_0^\infty g_2(R)V_2(R)dR} \right],$$

where $V$ is the sample volume, $f_p$ is the volume fraction of texture of each distribution in the sample, and $g(R)$ is the log-normal size distribution.

**Glass model in simulation.** Here, a 2D model glass system was considered. The model was developed by Shintani and Tanaka[48], in which a spherically symmetric Lennard–Jones (LJ) interaction were modified by including an anisotropic part. While the LJ interaction favours a crystalline state with long-range orders, the anisotropic part favours short-range five-fold symmetry. Briefly, the model potential has the following form:

$$U_{ij} = 4\epsilon\left(\left(\frac{\sigma}{r_{ij}}\right)^{12} - \left(\frac{\sigma}{r_{ij}}\right)^6 - \Delta\left(\frac{\sigma}{r_{ij}}\right)^6 f(\theta_i, \theta_j)\right).$$

Here, $f(\theta_i, \theta_j) = h(\theta_i - \theta_0)/\theta_c + h(\theta_i - \theta_0)/\theta_c - 64\theta_c/35$, where $h(x) = [1 -$

$x^2]^3$ for $-1 < x < 1$ and $h(x) = 0$ for $x \leq -1$ or $x \geq 1$. $\theta_i$ is an angle between the relative vector $\mathbf{r}_{ji} = \mathbf{r}_j - \mathbf{r}_i$ and the unit vector $\mathbf{u}_i$ that represents the orientation of the axis of particle $i$. $\theta_j$ is an angle between the relative vector $\mathbf{r}_{ij} = \mathbf{r}_i - \mathbf{r}_j$ and the unit vector $\mathbf{u}_j$.

**MC simulations.** Here, we performed standard MC simulations with the following Hamiltonian:

$$H = \sum_{\langle i,j \rangle} \left[ -J_1 (\mathbf{S}_i \cdot \mathbf{S}_j) + \mathbf{D}_{ij} \cdot (\mathbf{S}_i \times \mathbf{S}_j) \right] + \sum_{\langle\langle i,k \rangle\rangle} J_2 (\mathbf{S}_i \cdot \mathbf{S}_k) - B_z \sum_i S_{i,z}$$

where $\mathbf{S}_i$ is the spin on site $i$, $<i,j>$ and $<<i,k>>$ means the nearest neighbor (NN) and next-nearest neighbor (NNN), respectively. In our simulation, $|\mathbf{S}_i| = 1$ and a 2D amorphous system was employed. $J_1$ is the NN ferromagnetic coupling, and $J_2$ is the NNN antiferromagnetic coupling, while $\mathbf{D}_{ij} = D\hat{\mathbf{r}}_{ij}$ ($r_{ij} = |\mathbf{r}_{ij}|$ and $\hat{\mathbf{r}}_{ij} = \mathbf{r}_{ij}/r_{ij}$) is the vector of the DMI between NN sites $i$ and $j$. Here, NN is defined with a distance cutoff of 1.4, while NNN is defined as the particle which is not the NN but shares two NN with the center particle. During each MC step, we first select randomly a spin site, and then give it a random new spin direction, and finally use random number to determine the chance for whether to accept this new spin direction. To capture the physical properties in a broad $B_z$-$T$ phase plane, $B_z$ was sampled from 0.00 to 0.30 with interval of 0.01, $T$ was sampled from 0.0 to 3.0 with interval of 0.1 (Due to the limitation of MC simulation, $T=0.0$ was replaced by $T=0.01$). At each fixed $B_z$, the system was simulated by temperature scanning from high T to low T. And during MC simulations at each T, the system was first annealed for 25600×100000 MC steps, and following further relaxation of 25600×100000 MC steps for sampling.

**Calculation of topological charge.** The topological charge $Q$ is calculated according to its original definition:

$$Q = \frac{1}{4\pi}\int d^2r\, \mathbf{n}\cdot\left(\partial_x\mathbf{n}\times\partial_y\mathbf{n}\right),$$

where **n** is a unit vector describing the local spin direction. Since we are dealing with a discontinuous point set with amorphous positions, the Berg formula can be used. Thus, we first use Delaunay triangulation to completely split the entire plane into triangles. Then we calculated the integral in each triangle and sum them together. Mathematically, the integral in each triangle can be calculated as the solid angle ($\Omega_t$) of the three spins (at the three vertices of the triangle):

$$\exp\left(\frac{i\Omega_t}{2}\right) = \rho^{-1}\left[1 + \mathbf{S}_1\cdot\mathbf{S}_2 + \mathbf{S}_2\cdot\mathbf{S}_3 + \mathbf{S}_3\cdot\mathbf{S}_1 + i\mathbf{S}_1\cdot(\mathbf{S}_2\times\mathbf{S}_3)\right],$$

$$\rho = \left[2(1+\mathbf{S}_1\cdot\mathbf{S}_2)(1+\mathbf{S}_2\cdot\mathbf{S}_3)(1+\mathbf{S}_3\cdot\mathbf{S}_1)\right]^{\frac{1}{2}},$$

$$Q = \frac{1}{4\pi}\sum_t \Omega_t.$$

## Acknowledgements

This research was supported by the Strategic Priority Research Program of the Chinese Academy of Sciences (Grant No. XDB30000000), National Key Research and Development Plan (Grant No. 2018YFA0703603), National Natural Science Foundation of China (Grant No. 52192600, No. 52130108, No. 11790291, No. 61999102, No. 61888102, No. 51871234 and No. 51971238) and Natural Science Foundation of Guangdong Province (Grant No. 2019B030302010).

## Author contributions

H. B. proposed the study and supervised it with Z. L.; W. W., J. L., L. S., H. J., and L.G. prepared the samples and performed XRD and HRTEM experiments; W. W. performed the magnetic characterization and magnetotransport experiments, and analyzed the data with H. Z., H.W., Z. L., and H. B.; W. W., H. Z., Z. F., D.M., and H. B. applied the beam time for (SANS); W. W. and Z. L. prepared the sample of SANS while H. J., C. H., Y. K., Y. S., and K. H. performed SANS experiment; W. W., H. Z., C. C., Z. F., D. M., and H. B. analyzed the SANS data; H. W. performed the Lorentz transmission electron microscopy, and analyzed them with W. W., H. Z., Z. L., and H. B.; H. Z. and M. L. developed the Monte Carlo simulations, and discussed with W. W. and H. B.; W. W., H. Z., H. W., Z. L. and H. B. wrote the manuscript with input and comments from all authors.

## Competing interests

The authors declare no competing interests.

## Data availability

The data that support the findings of this study are available from the corresponding author upon reasonable request.